\documentclass[preprint2]{aastex}
\usepackage{amsmath}
\usepackage{natbib}
\usepackage{ulem}
\usepackage{lineno}

\usepackage{color}

\shorttitle{Self-absorption in the solar TR} \shortauthors{Yan et al.}

\begin{document}


\title{Self-absorption in the solar transition region}

\author{Limei Yan\altaffilmark{1,2}, Hardi Peter\altaffilmark{2}, Jiansen He\altaffilmark{1}, Hui Tian\altaffilmark{3}, Lidong Xia\altaffilmark{4}, Linghua Wang\altaffilmark{1}, Chuanyi Tu\altaffilmark{1}, Lei Zhang\altaffilmark{1}, Feng Chen\altaffilmark{2}, Krzysztof Barczynski\altaffilmark{2}}

\altaffiltext{1}{School of Earth and Space Sciences, Peking University, 100871 Beijing, China, jshept@gmail.com} \altaffiltext{2}{Max Plank Institut f\"ur Sonnensystemforschung, \\Justus-von-Liebig-Weg 3, 37077 G\"ottingen, Germany} \altaffiltext{3}{Harvard-Smithsonian Center for Astrophysics, 60
Garden Street, Cambridge, MA 02138, USA} \altaffiltext{4}{Shandong Provincial Key Laboratory of Optical Astronomy and Solar-Terrestrial
Environment, School of Space Science and Physics, Shandong University, Weihai, China}

\begin{abstract}
Transient brightenings in the transition region of the Sun have been studied for decades and are usually related to magnetic reconnection. Recently, absorption features due to chromospheric lines have been identified in transition region emission lines raising the question of the thermal stratification during such reconnection events.
We analyse data from the Interface Region Imaging Spectrograph (IRIS) in an emerging active region. Here the spectral profiles show clear self-absorption features in the transition region lines of Si\,{\sc{iv}}. While some indications existed that opacity effects might play some role in strong transition region lines, self-absorption has not been observed before.
We show why previous instruments could not observe such self-absorption features, and discuss some implications of this observation for the corresponding structure of reconnection events in the atmosphere. Based on this we speculate that a range of phenomena, such as explosive events, blinkers or Ellerman bombs, are just different aspects of the same reconnection event occurring at different heights in the atmosphere.

\end{abstract}


\section{Introduction}

In general, the emission originating in the transition region
between the chromosphere and the corona as well as from the corona itself is optically thin for emission lines observed on the Sun. Mostly, the low density of the plasma in the upper atmosphere allows each emitted photon to escape without further scattering. There are some exceptions, though. In particular, strong resonance lines have been reported to show effects of optical depth through the analysis of line  ratios. The resonance lines of Li-like ions, e.g. C\,{\sc{iv}} at 1548\,\AA\ and 1550\,\AA, should have a line ratio of about 2 under optically thin conditions (according to their Einstein $A$ coefficients). However, when observed towards the limb of the Sun, implying a longer line-of-sight, an increased non-thermal width and a reduced line ratio
indicate that optical depth effects become (marginally) important \citep{Mariska:1992,Peter:1999}. Despite this indirect evidence, no self-absorption features have been found in lines forming in the transition region or at higher temperatures.

If such self-absorption features could be found, they would give important information on the structure of the solar atmosphere. This is in analogy  to the arguments based on the absence of absorption lines in the extreme ultraviolet (EUV) originating from the chromosphere.
While dynamic models of the photosphere and chromosphere \citep{Carlsson+Stein:1997} overall reproduce the details of observed chromospheric spectra, they have been criticized
because the average background atmosphere with a monotonic decreasing temperature should produce EUV absorption lines \citep{Kalkofen:2012}. Concerning the transition region emission, the existence or absence of self-absorption features is of interest to understand the thermal structure of the atmosphere along the line of sight. Traditionally, the transition region was considered as a thin layer separating the chromosphere and the corona. However, recent observation of pockets of hot gas in the cool chromosphere questioned this paradigm \citep{Peter2014Sci}, and the observation of so far unreported self-absorption features could hold the key to better understand the deviation of the realistic transition region from our traditional knowledge.

A whole zoo of phenomena have been identified on the Sun that show themselves through  peculiarities of transition region emission lines in the EUV, either by enhanced emission, changes in the spectral profiles, or both.
Among them are explosive events \citep{Dere1989SoPh,Innes1997Natur}, blinkers \citep{Harrison+al:1999,Peter+Brkovic:2003}, and the recently found Ellerman-bomb-type events \citep{Peter2014Sci}.

The physical nature of these events is not yet fully clear. The common feature is that all these events, just as the traditional Ellerman bombs \citep{Ellerman:1917,Georgoulis:2002}, are thought to be the response to a reconnection event. Should this be the case, the main reason for the quite different appearance in EUV and visible light might be mainly due to the exact location of the reconnection site. If it is located very deep in the atmosphere, it will be visible only in emission from the chromosphere, e.g., H$\alpha$, as traditional Ellerman bombs \citep{Schmieder:2004}. On the other end of the scale, if the reconnection happens higher up, it will be visible as an explosive event, mostly in transition region lines, with only a small counterpart in the chromosphere \citep{Brkovic+Peter:2003,Doyle:2005}.

Here we present observations from the Interface Region Imaging Spectrograph
\citep[IRIS,][]{DePontieu2014SoPh} that share spectral features with previous EUV brightenings, but for the first time show a self-absorption feature in a transition region line.

\section{Observation}\label{S:obs}
The event we concentrate on in this study was observed by IRIS from 09:20 UT to 09:26 UT on 2013 December 6, and showed a peculiar self-absorption feature in Si\,{\sc{iv}}. During this campaign, IRIS was observing an emerging active region
in sit-and-stare mode with an exposure time of 4~s. The spatial sampling  of the spectrograph (along the $0.35''$ wide slit) is  $0.167''$ per pixel. In the spectral direction an on-board two-pixel binning was applied leading to a spectral sampling of 26~m\AA\ per pixel, corresponding to about 5.5\,km\,s$^{-1}$ per pixel in Doppler units. Considering the spectral resolution of about 30\,m\AA\ in the FUV \citep{DePontieu2014SoPh,Tian2014Sci}, in our dataset the spectral resolution is governed solely by the spectral sampling.

The main emphasis is on the analysis of the profiles of EUV emission lines from the transition region, in particular of the two resonance lines of Si\,{\sc{iv}} at 1393.76\,\AA\ and 1402.77\,\AA, the  C\,{\sc{ii}} lines at 1334.54\,\AA\ and 1335.71\,\AA, and the forbidden lines of O\,{\sc{iv}} at 1399.78\,\AA\ and 1401.16\,\AA.
Besides this, we also investigated the IRIS slit-jaw images in the 1400~\AA\ band. This shows a mixture of emission from the chromosphere (Lyman continuum of Si\,{\sc{i}}) and the transition region (Si\,{\sc{iv}} lines). These slit-jaw images were taken with an
exposure time of 4~s and a cadence of ~16~s. Their spatial sampling is $0.167''$ per pixel.
We use the IRIS level~2 data which are publicly available at http://iris.lmsal.com.

To investigate the signatures of this event at different temperatures we also examined  images taken by the Atmospheric Imaging Assembly
\citep[AIA,][]{Lemen2012SoPh} onboard the Solar Dynamics Observatory. Their spatial sampling is $0.6''$ per pixel. Observations in the ultraviolet (UV)
passbands (1600\,\AA\ and 1700\,\AA) have a cadence of ~24~s, while the extreme ultraviolet (EUV) observations (304\,\AA, 171\,\AA, 193\,\AA,
211\,\AA, 335\,\AA, 131\,\AA, and 94\,\AA) have a cadence of 12~s. We use the longitudinal magnetograms obtained by the Helioseismic and
Magnetic Imager \citep[HMI,][]{Schou2012SoPh} to study the evolution of the corresponding magnetic field. These magnetograms have a
spatial sampling of $0.5''$ per pixel and a cadence of 45~s.

We align the IRIS spectra and slit-jaw images using the fiducial marks crossing the slit. The AIA 1600\,\AA\ images are used to achieve a coalignment between the IRIS and AIA
data. The HMI images are aligned to AIA 1600\,\AA\ images using a cross-correlation technique.

\section{Transition region brightening with \\(self-)absorption}\label{S:TR.brightening}

\begin{figure*}
\centering
\includegraphics[width=14cm]{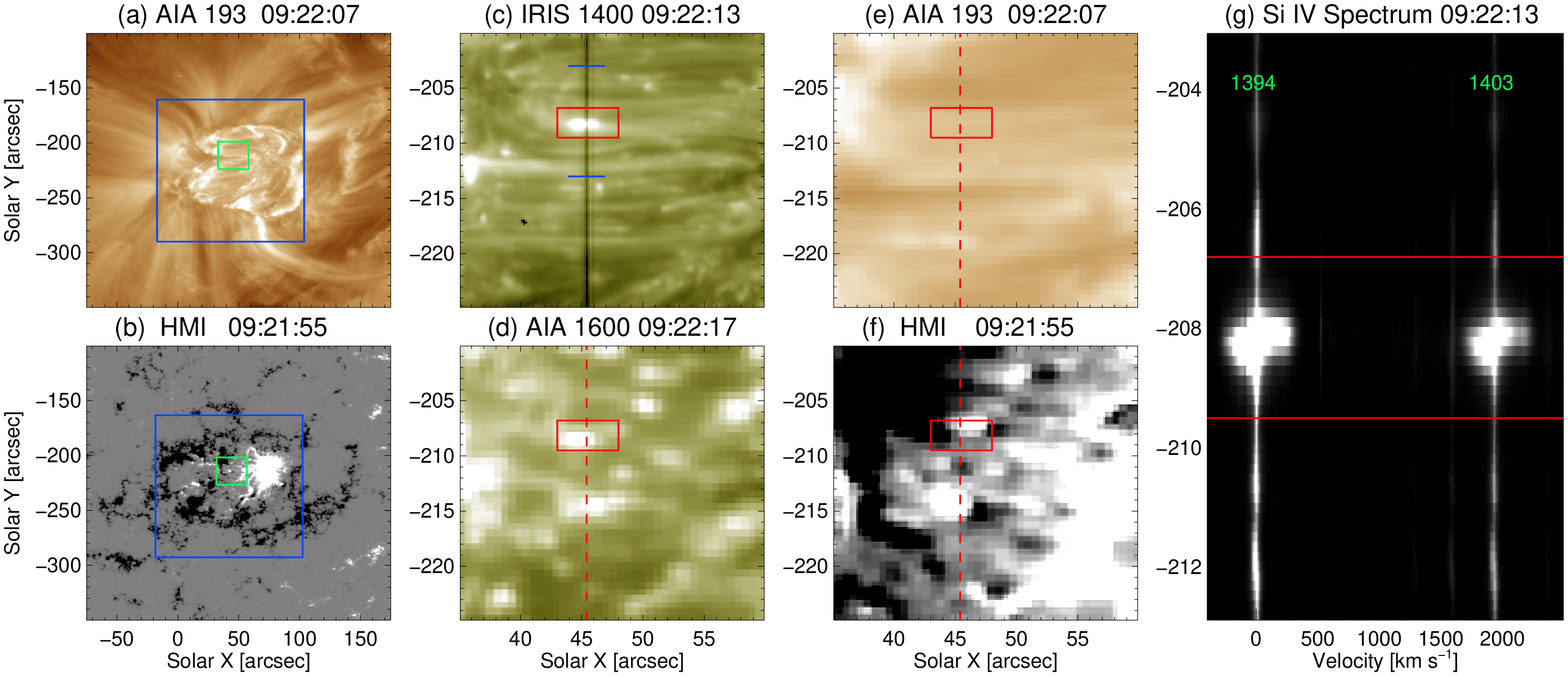}
\caption{Overview of the the active region under study during the time when self-absorption features in Si\,{\sc iv} are observed.
\newline
(a): AIA 193 \AA\ showing 1.5\,MK hot plasma. (b): HMI longitudinal magnetic field.
The blue boxes in (a) and (b) show the full FOV of IRIS, the green boxes indicate the subfield shown in panels (c) to (f). (c): IRIS slit-jaw-image at 1400 \AA\ mostly showing 10$^5$\,K plasma. The vertical black line is the spectrograph slit. (d): AIA 1600\,\AA\ showing chromospheric and transition region emission. (e) and (f): Zoom into panels (a) and (b). The red box in (c)-(f) highlights the position of the event we concentrate on in this study.
The vertical red dashed lines in (d)-(f) indicate the location of the slit. (g): The spectrum covering both Si\,{\sc iv} lines at 1394\,\AA\ and 1403\,\AA. The wavelength is given in Doppler  velocity with respect to Si\,{\sc iv} 1394\,\AA. The two horizontal red lines are at the same Y position as the red boxes in (c)-(f). The two blue lines crossing the slit in panel (c) indicate the extension of the spectrum shown in panel (g) along the slit.
See Sect.\,\ref{S:TR.brightening}. A movie is available from online materials.} \label{fig1}
\end{figure*}

The 1400 \AA\ slit-jaw images show a bright point appearing just to the left
of the IRIS slit at 09:20:25~UT. It grew and extended all the way to the right side of the slit around 09:22:13~UT (Figure~\ref{fig1}) and disappeared around 09:24:02. These slit-jaw images show that the IRIS slit was placed across this brightening event so that the EUV spectra will show (part of) its evolution. For context, the AIA 193 \AA\ image showing the coronal 1.5\,MK plasma and the (longitudinal) magnetic field of the whole active region are displayed in Figure~\ref{fig1}, too. The event can also be identified with in AIA 1600 \AA\ images, with an elliptical shape similar to the IRIS 1400\,\AA\ images. However, the event is not seen in the 1700\,\AA\ images, suggesting that the response in 1600 \AA\ is due to the emission of C\,{\sc iv} at 1548\,\AA\ and 1550\,\AA\ rather than due the UV continuum.

During the event, the intensity of Si\,{\sc iv} increases roughly by a factor of 20 compared to the average emission  before, which is consistent with the 1600\,\AA\ channel being dominated by C\,{\sc iv}. This also implies that the 1400\,\AA\ channel is dominated by emission from about 0.1\,MK hot plasma, too, in this case.
The maximum response of 171 \AA\ channel of AIA \citep[$10^{-24}$ DN\,cm\,s$^{5}$\,s$^{-1}$\,pix$^{-1}$ at just below 10$^6$\,K;][]{Boerner+al:2012} is $10^3$ times higher than at the formation temperature of Si\,{\sc iv}, so if the plasma in the current event is heated only to some  10$^5$\,K, no counterpart in the AIA 171 \AA\ channel, or channels showing even hotter plasma is expected. Observing no signature of the event in 171\,\AA\ or 193\,\AA\ (see Figure~\ref{fig1})
is consistent with the event not producing any plasma of 1\,MK or hotter (at least not in significant abundance).

Inspection of the HMI\ magnetogram in Figure~\ref{fig1} shows that the event occurred in a mixed-polarity region. We find
no obvious change of the magnetic flux magnitude directly at the site of the event. However, the flux of the negative polarity to the upper right of the event
(X: $47'' - 51''$, Y: $-208'' - 206''$, see red box in Figure~\ref{fig1}f)  shows a significant decrease. This  might be related to flux cancellation, and depending on the magnetic structure, this flux cancellation might be directly connected to the event we observe.

\subsection{Tilted spectra}
During the event, the line profiles of the Si\,{\sc iv} lines and the C\,{\sc ii} lines become
very broad due to the strong enhancement in the red and blue wings (Figure~\ref{fig2}). A similar but less strong enhancement is also present in the Mg\,{\sc ii} lines at 2796.347 \AA\  and 2803.523 \AA. This indicates that the event is covering also the uppermost tip of the chromosphere. In contrast, no significant enhancement is found in the O\,{\sc iv} 1400\,\AA\ and 1401\,\AA\ forming just below 0.2\,MK. This underlines that the event does not reach coronal temperatures and might be limited to temperatures below $\sim\,0.1$\,MK. In contrast to the bombs studied by \citet{Peter2014Sci}, here the forbidden O\,{\sc{iv}} 1401\,\AA\ line is visible throughout the whole brightening (cf.\ middle row of Figure \ref{fig2}).

The line profiles of the Si\,{\sc iv} lines show a spectral tilt  (Figure~\ref{fig1}g): the profiles have dominant red wings in the southern part of the brightening (y = $-208.385''$), they become roughly symmetric in the middle of the brightening (y = $-208.219''$) and finally show dominant blue wings in the northern part (y = $-208.052''$). These type of spectra are well known and are typically associated with twisting \citep{DePontieu+al:2014} or rotation \citep{Curdt2011A&A} of the plasma in helical structures \citep{Li+al:2014}.
The C\,{\sc ii} and Mg\,{\sc ii} line profiles also show a similar feature.

The strong enhancement of the line and the wings is very clear in the profiles shown in Figure~\ref{fig2}. During the peak time of the brightening, the line profile shows extended wings reaching out to Doppler shifts of more than 200\,km\,s$^{-1}$ in the blue and red.
Moreover, there is a clearly distinguishable component redshifted by $\sim$250\,km\,s$^{-1}$ in the phase of highest Si\,{\sc{iv}} emission (light blue profile; Fig.\,\ref{fig2}, top and middle row). This and the clear excess of emission in the wings throughout the event suggests the presence of non-resolved plasma flows with a wide range of velocities reaching speeds (along the line of sight) in excess of 200\,km\,s$^{-1}$ in these events.
These speeds are highly supersonic and will be comparable or even higher than the Alfv\'en speed (depending on the assumptions for the magnetic field and density at the location of the event). Such high velocity components have not been observed in the spectra of explosive events or blinkers that have been mostly observed in quiet Sun network areas, and they are even higher than the speeds found in Ellerman-bomb-type events recently found in Si\,{\sc
iv} \citep{Peter2014Sci}. This underlines that the event we deal with here is a violent disruption of the upper chromosphere and transition region.

\subsection{Chromospheric absorption lines}\label{S:abs.lines}

Embedded in the emission line profiles of Si\,{\sc iv} and C\,{\sc ii} clear signatures of chromospheric absorption lines are present during the strong brightening. This is illustrated in Figure~\ref{fig2}, where the rest wavelength of several lines of Ni\,{\sc ii} (1335.203 \AA, 1393.33
\AA) and Fe\,{\sc ii} (1392.817 \AA, Fe\,{\sc ii} 1393.589 \AA)
are indicated.
These chromospheric absorption lines  in the emission line profiles of a transition region line have been first detected by  \citet{Peter2014Sci}. A Gaussian fit to the Ni\,{\sc ii} absorption feature in the Si\,{\sc iv} 1394\,\AA\ line gives a full width at half maximum ({\sc{fwhm}}) of about 11\,km\,s$^{-1}$ (see sample spectrum in Figure\,\ref{fig3}a), which corresponds to two (binned) pixels in the spatial direction (cf.\ Sect.\,\ref{S:obs}). The thermal width of Ni\,{\sc ii} would be well below 5 km\,s$^{-1}$. Thus the width of the narrow
chromospheric absorption features will be mainly  determined by the instrument. This is consistent with the Ni\,{\sc ii} absorption features forming in a cool layer {\itshape above} the hotter formation region of Si\,{\sc iv}.

Because the data set investigated by
\citet{Peter2014Sci} was a raster map, they could not study the temporal evolution of the chromospheric lines. Here we see that at no time during the evolution of the event these chromospheric lines appear in emission (see in particular the top left panel of Figure~\ref{fig2}). Only when the Si\,{\sc iv} and  C\,{\sc ii} lines get strong enough in their wings while brightening and broadening, these chromospheric lines show themselves in absorption. This result will be of interest when investigating new models of these events, because they contain information on the timing of the event that might be crucial to distinguish between different model scenarios. For example, \citet{Peter2014Sci} interpreted the absorption spectra as being due to a (shallow) layer of chromospheric material that exists above a reconnection site in the lower chromosphere where Si\,{\sc{iv}} originates. The fact that the chromospheric lines are not seen in emission, but in absorption might be an indication that the above-lying shallow chromospheric layer might be unaffected by the reconnection event. However, detailed modeling including radiative transfer will be needed to draw conclusions on this.

The Ni\,{\sc ii} 1393.33 \AA\  line has been observed in bursty explosive events \citep{Ning2004A&A}. In contrast to our observation and to  \citet{Peter2014Sci}, there the line was always seen in emission. That Ni\,{\sc ii} is seen in emission may be understood if the explosive event occurs in the upper chromosphere,
as suggested by \cite{Chen2006SoPh}. They proposed that  implosive bursty transition region explosive events could occur only
under two necessary conditions, the modulation of p-mode
oscillations and that the reconnection site is located in the
upper chromosphere. The event we study here, however, is in all probability occurring at lower heights.

We can estimate the electron density $n$ through the intensity ratio of Si\,{\sc iv} 1403
\AA\  and O\,{\sc iv}
1401 \AA\ \citep{Peter2014Sci}, and find $\log{n}\,[\mbox{cm}^{-3}]$ to be about 12.2 to 12.5 (under the assumption of ionization equilibrium). In contrast to  \cite{Peter2014Sci}, in our observation both  O\,{\sc iv} lines at 1400\,\AA\  and 1401\,\AA\ are present, so that we can also use these two forbidden lines for density diagnostics and find $\log{n}\,[\mbox{cm}^{-3}]$ to be about 11. From this we conclude that the electron density in the source region of
 our event is in the range of  $10^{11}$\,cm$^{-3}$ to $10^{12}$\,cm$^{-3}$, which are chromospheric
densities. This result, together with the presence of the absorption lines of
Ni\,{\sc ii} and Fe\,{\sc ii}, suggests that our event probably occurs in the
middle chromosphere. This might suggest that one of the main differences between an explosive event and the event we report here or the Ellerman-bomb-type events by  \cite{Peter2014Sci}
might just be the height in the atmosphere where the reconnection occurs.

\subsection{Self-absorption of Si\,{\sc iv}}\label{S:self.abs}

The most peculiar spectral feature in the event we investigate here is the clear presence of self-absorption in a transition region line that has not been reported before. This feature is seen at both  Si\,{\sc iv} lines at 1394\,\AA\ and 1403\,\AA.

\begin{figure*}
\centering
\includegraphics[width=14cm]{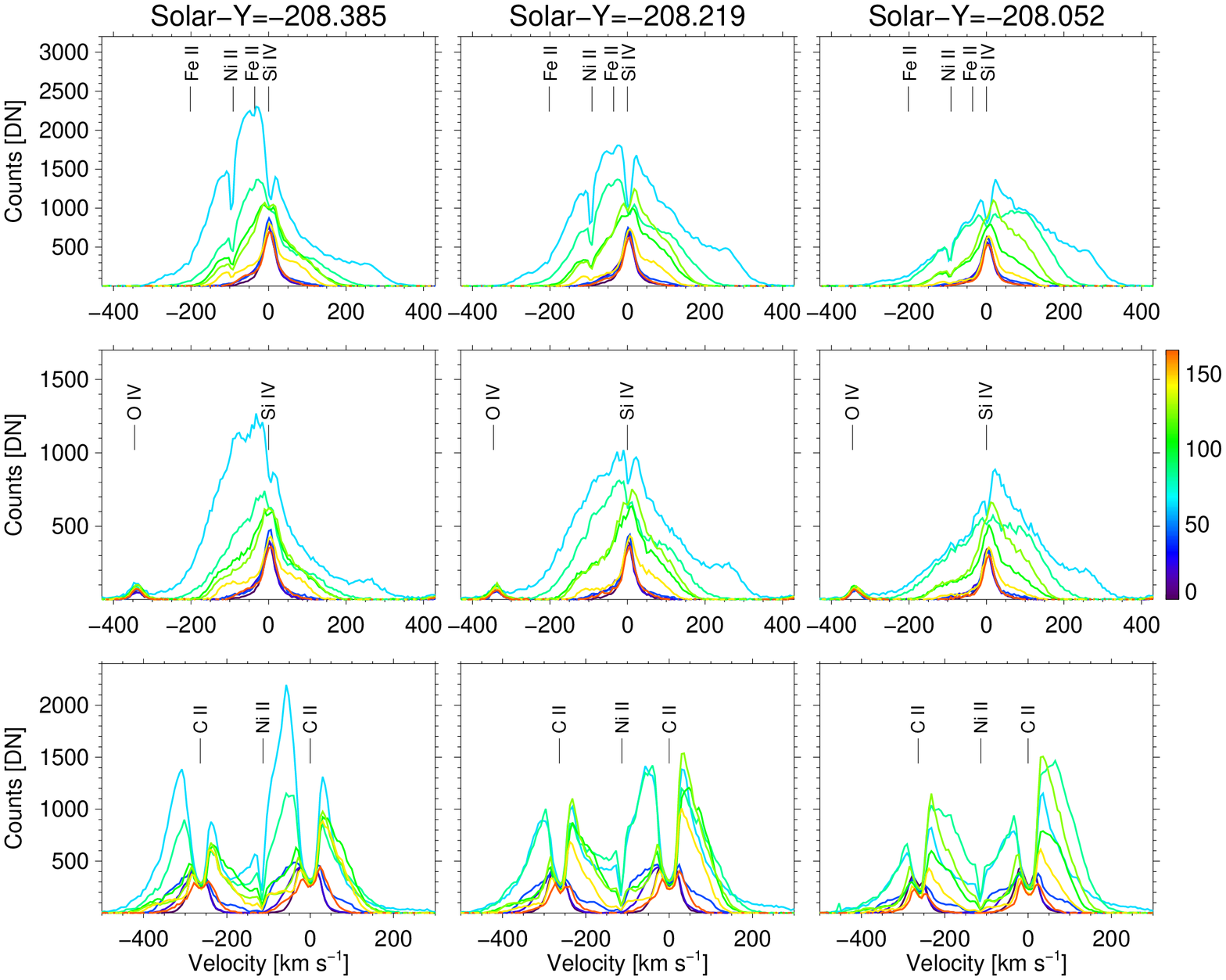}
\caption{Spectra covering Si\,{\sc iv} and C\,{\sc ii} lines as a function of space and time.
\newline
The three columns show the spectra at three adjacent spatial pixels (with the solar Y location in arcsec) in the middle of the red boxes in Figure\,\ref{fig1}(c)--(f).
The three rows show spectral windows centered around Si\,{\sc iv} 1394\,\AA\ (top), Si\,{\sc iv} 1403\,\AA\ (middle), and C\,{\sc ii} 1335\,\AA\ and 1336\,\AA\ (bottom). All wavelength are given in Doppler shift units.
Each panel shows the time evolution of the line profiles over 3\,min starting at  09:21:16.
Different colors represent different times (according to the color bar showing time in s).
Besides the rest wavelengths of the main lines, also O\,{\sc{iv}} and some chromospheric absorption lines of Ni\,{\sc ii} and Fe\,{\sc ii} are labeled. See Sects.\,\ref{S:abs.lines} and \ref{S:self.abs}.} \label{fig2}
\end{figure*}

In the evolution of the line profiles of both  Si\,{\sc iv}
lines in Figure\,\ref{fig2} near the rest wavelength of the respective  Si\,{\sc iv}
line (indicated by the short vertical lines) a narrow absorption feature appears.

The dips reach a depth of up to 1/3  (sometimes even  1/2) of the
peak emission (cf. Figures \ref{fig2} and \,\ref{fig3}).
A Gaussian fit to these dips gives a width of about 17\,km\,s$^{-1}$({\sc{fwhm}}; cf.\ Figure\,\ref{fig3}a),
which is clearly resolved by the IRIS spectrometer and significantly broader
than the  chromospheric absorption lines discussed in Sect.\,\ref{S:abs.lines}. Also molecular absorption
lines would not be resolved by IRIS because they form at very cool
temperatures only \citep{Schmit2014A&A}, and thus would show line widths comparable to the chromopsheric absorption features in our data set. This indicates that the absorption features
 near the  Si\,{\sc iv} rest wavelength are not due to chromospheric or molecular lines. Furthermore a search of lines lists did not reveal obvious potential chromospheric
lines that could lead to these absorption features at the rest wavelengths
of  Si\,{\sc iv}. If the absorption would be due to molecules one would expect
not only one line, but a whole band of lines as reported by \citet{Schmit2014A&A}.
Also the simple fact that in both  lines the absorption feature appears at the rest wavelength of Si\,{\sc iv} (within the errors) indicates that this is an absorption feature related to  Si\,{\sc iv} itself.

The narrow dip in the profile of Si\,{\sc iv}  at its rest wavelength is also {\itshape not} due to a bi-directional or multi-component flow. If that would be the case, then one would expect the line profile to be composed of two or more Gaussians at different Doppler shifts representing the different flow components. To mimic the dip as observed here with  a width of about 17 km/s and a depth of 1/3 of the peak emission, one has to add two  Gaussians with a {\sc{fwhm}}  of 27\,km\,s$^{-1}$ separated by 35\,km\,s$^{-1}$in wavelength. However, this composite profile would have a  {\sc{fwhm}} of only 60\,km\,s$^{-1}$, while the observed profiles have a {\sc{fwhm}} of 150\,km\,s$^{-1}$ to 300\,km\,s$^{-1}$. If one would assume two Gaussians wide enough to match these large line widths, then also the central dip would be very broad, in contrast to observations. The only option would be to assume that the profile is composed by a range of individual narrow Gaussians with Doppler shifts distributed from some $-150$\,km\,s$^{-1}$ to some $+150$\,km\,s$^{-1}$ with no components just at zero Doppler shift --- which is quite unlikely, in particular because this very special line-of-sight arrangement would have to be found in many cases. Most important, the dips in the Si\,{\sc iv} lines at 1394\,\AA\ and 1403\,\AA\ show different depths, which is not expected if the dip would come about by a composition of flows under optically thin conditions. Thus one can safely exclude the possibility that the dip at the rest wavelength of Si\,{\sc{iv}} is caused by a bi- or multi-directional flow.

Thus one ends up with the most likely conclusion that the features seen at the rest wavelength at both Si\,{\sc iv} lines have to be due to self-absorption. Major support for this interpretation is given by the ratio of the two Si\,{\sc iv} lines at 1394\,\AA\ and 1403\,\AA. Under optically thin conditions their line ratio should equal the ratio of the oscillator strength, which is about 2 for these Li-like resonance lines \citep[see also discussion in][]{Peter2014Sci}.
During our event we see a decrease of the Si\,{\sc iv} ratio from about 2 to approximately 1.7. After the event the ratio is about 2 again, showing that opacity effects become important during the brightening.
Because the peak count rates are 1000 and higher (for the 1394\,\AA\ line, cf. Fig.\ \ref{fig2}), the uncertainty is of the order of a few percent \citep[for the error estimate see] []{DePontieu2014SoPh}. The uncertainty for the integrated line intensity is even smaller, and we estimate the relative error for the line ratio to be some 3\%. This underlines that the variation of the line ratio we see is significant.


The opacity effects should be more obvious in the stronger of the two Si\,{\sc iv} lines, i.e. the one at 1394\,\AA. This is indeed found in our data, where the dip at the rest wavelength is always stronger in the 1394\,\AA\ line (Figure\,\ref{fig2}, top row) than in the 1403\,\AA\ line (Figure\,\ref{fig2}, middle row).%
\footnote{%
The double-peaked Si\,{\sc{iv}} profile shown in \cite{Peter2014Sci} is due to an optically thin bi-directional flow. However, their profile was quite different in that the width of the central dip is very broad, the ratio of the 1394\,\AA\ line to the 1403\,\AA\ line is close to 2, and that the dip has the same depth in both lines.
}

Even though we concentrate on one event only in this study, this self-absorption feature is not a singular phenomenon. Just to show that this occurs more frequently, we show another example in Figure~\ref{fig3}c. This sample spectrum is taken in the same active region, just a bit further South (c.\ Figure\,\ref{fig3}b).

\section{Absorption features with SUMER resolution}\label{S:sumer}

Why have these chromospheric absorption lines and the Si\,{\sc iv} self-absorption not been observed with previous instruments? In particular, could these absorption features have been observed
with SUMER? Before IRIS, SUMER  \citep{Wilhelm95} and HRTS \citep{Brueckner1983ApJ} were the instruments with the best spectral and spatial resolution for EUV spectroscopy, roughly equal in performance. To answer the above question, we perform a simple experiment degrading the IRIS spectra to SUMER characteristics and check for the presence of (self-)absorption features.

The exposure times of SUMER and IRIS are comparable for dynamics studies, e.g.,\ the explosive event study by \cite{Innes1997Natur} used 5\,s exposures, while here we have 4\,s exposures.
However, the
spatial resolution
(and the sampling) of IRIS is better by about a factor of almost six; the spatial sampling of IRIS being 0.167$''$/pixel along the slit, and for SUMER only 1$''$/pixel. Likewise the spectral resolution of IRIS is better by about a factor of three. In the following we adopt the widths for the  instrumental spectral profile ({\sc{fwhm}}) of 31.8~m\AA\ for
IRIS \citep{Tian2014Sci} and 85.8~m\AA\ for SUMER \citep{Chae1998ApJ}.

To simulate a SUMER spectrum from IRIS data, we take six adjacent IRIS profiles to account for the difference in spatial
sampling, deconvolve the spectra for the instrumental broadening, add them up to give one SUMER spatial pixel and convolve the resulting profile with the SUMER instrumental spectral profile. This would mimic an observation with the SUMER 0.3$''$ slit, roughly matching the 0.35$''$ wide IRIS slit.

The IRIS profiles of Si\,{\sc iv} 1394\,\AA\ at six adjacent spatial pixels during the event showing (self-) absorption are shown in Figure~\ref{fig4}a\,to\,f. Because the event is compact (smaller than 1$''$) not all the profiles show the absorbtion features. For the spectral deconvolution we use a maximum entropy code.%
\footnote{Maximum entropy image deconvolution code ``mem96.pro'' by Jongchul Chae, \\http://www.bbso.njit.edu/{$\sim$}chae/IDL/idl.html}
The average of the six IRIS profiles without and with deconvolution are shown in panels g and h of Figure\,\ref{fig4}. They appear to be quite similar. The final spectrum convolved with the SUMER\ spectral profile and rebinned to the SUMER spectral sampling is shown in Figure\,\ref{fig4}i. This  can be
considered as the profile  SUMER would have observed from this event.

From this experiment it is clear that SUMER would not have been capable to observe the chromospheric absorption or the Si\,{\sc iv} self-absorption. This is not only because of the spatial resolution, but mainly because of the limited spectral resolution. In a more real situation the small features still visible at the location of the (self-)absorption in Figure\,\ref{fig4}i would not have been visible with SUMER.  Typically, for dynamic studies the 1$''$ slit would have been used\footnote{The 0.3$''$ slit was used with SUMER mainly for observations  in very bright lines, e.g. Ly-$\alpha$, to keep the count rates at an acceptable level.}, so that one would have to deteriorate the spatial sampling by another factor three. This would then further smooth the line profile to be observed with SUMER, because the feature is smaller than 1$''$. Because our IRIS observations were acquired in a sit-and-stare mode, we do not have a raster map available and cannot take this experiment that far. Still we can expect that the weak remnants of the absorption and self-absorption features would be further diminished and likely
would be completely absent in real SUMER observations.

\section{Discussion and conclusions}\label{S:discussion}

In recent observations the tilt of spectral lines has been attributed to torsional or rotaional motions \citep{Curdt2011A&A,Curdt2012,DePontieu+al:2014,Li+al:2014}. However, the velocities we observe here are exceeding
100\,km\,s$^{-1}$, sometimes up to 250\,km\,s$^{-1}$, hence significantly larger than reported before. That these velocities get close to or even exceed the the Alfv\'en speed questions the interpretation for a torsional motion in our event. Instead it suggests that this outflow has the nature of a bidirectional jet. This is the traditional interpretation for transition region explosive events in the quiet Sun  \citep[e.g.,][]{Dere1991JGR,Innes1997Natur} and also for the recently reported Ellerman-bomb-type events in transition region lines \citep{Peter2014Sci}.
Thus we assume that at the basis of the event we report here, some reconnection is causing strong reconnection outflows which causes the strong brightening
and broadening of the Si\,{\sc{iv}} profile.

The interpretation of the event reported here is based on the presence of chromospheric absorption lines and self-absorption features in transition region lines. We showed that such (self-)absorption features were not observable with previous EUV\ spectrographs because of instrumental limitations, in particular in terms of spectral resolution.

In Figure\,\ref{fig5} we give a cartoon summary of a possible scenario.
At a height where in an equilibrium model the middle chromosphere would be located, a reconnection event gives rise the increase in intensity of Si\,{\sc{iv}} through heating of the previously cool chromospheric plasma. At the same time the multiple flows created by the reconnection cause a very strong broadening (panel 1 of Figure\,\ref{fig5}).

Above the location of the heating and the origin of the strong Si\,{\sc{iv}} emission the pre-existing upper atmosphere is still present. The thin chromospheric layer above will give rise to the absorption lines of Ni\,{\sc ii} and Fe\,{\sc ii} (panel 2 of Figure\,\ref{fig5}).
Actually, the chromospheric absorption features and the high density derived from line ratios are the main arguments to place the origin of the strong Si\,{\sc{iv}} emission in the middle chromosphere in the first place. This chain of argument is following  \cite{Peter2014Sci}.

\begin{figure*}

   \centering
    \includegraphics[width=14cm]{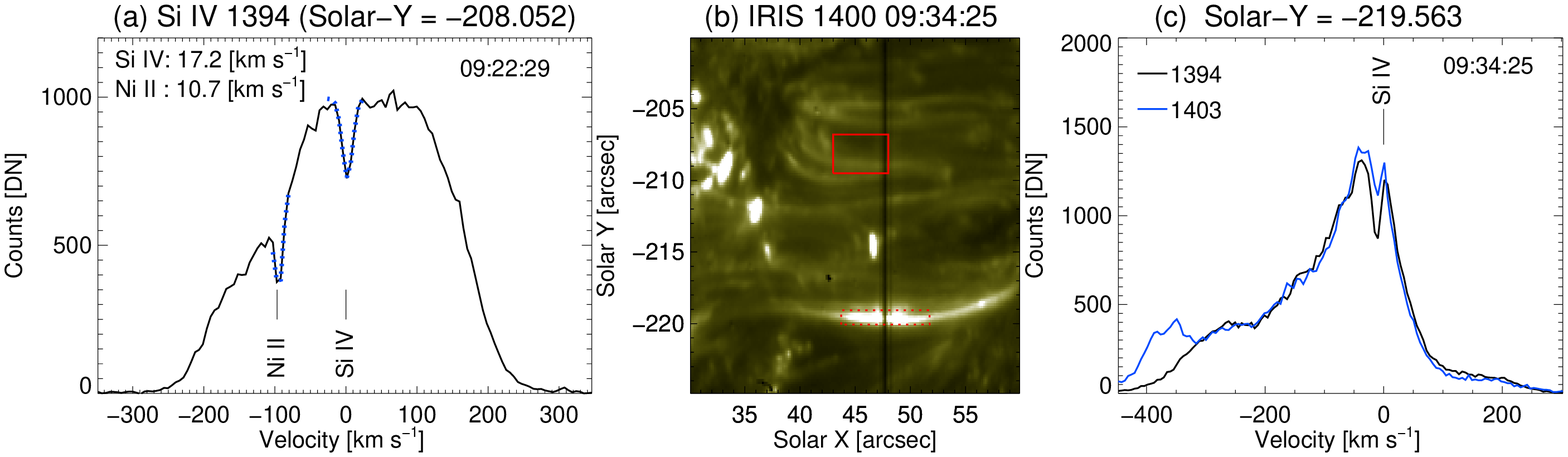}

\caption{Self-absorption of Si\,{\sc iv} in different features on the Sun.
\newline
(a): Example of the fitting results of the (self-)absorption features in
Si\,{\sc iv}. The blue dotted curves show (inverse) Gaussians fitted to the
absorption features. The numbers in the top left refer to the width (FWHM)
of the Ni\,{\sc ii} absorption and the Si\,{\sc iv} self-absorption.
(b): IRIS slit-jaw-image at 1400\,\AA\ obtained at 09:34:25~UT. The red boxes
indicate the positions of the profiles shown in panels (a) and (c). The respective
regions show different characteristics, but profiles from both show the self-absorption
features. (c): Sample profiles of both Si\,{\sc iv} lines at 1394\,\AA\ and
1403\,\AA\ from a loop top showing self-absorption features at the site of
a cool loop.  See Sects.\,\ref{S:abs.lines} and \ref{S:self.abs}.}
      \label{fig3}
\end{figure*}

\begin{figure*}
\centering
\includegraphics[width=14cm]{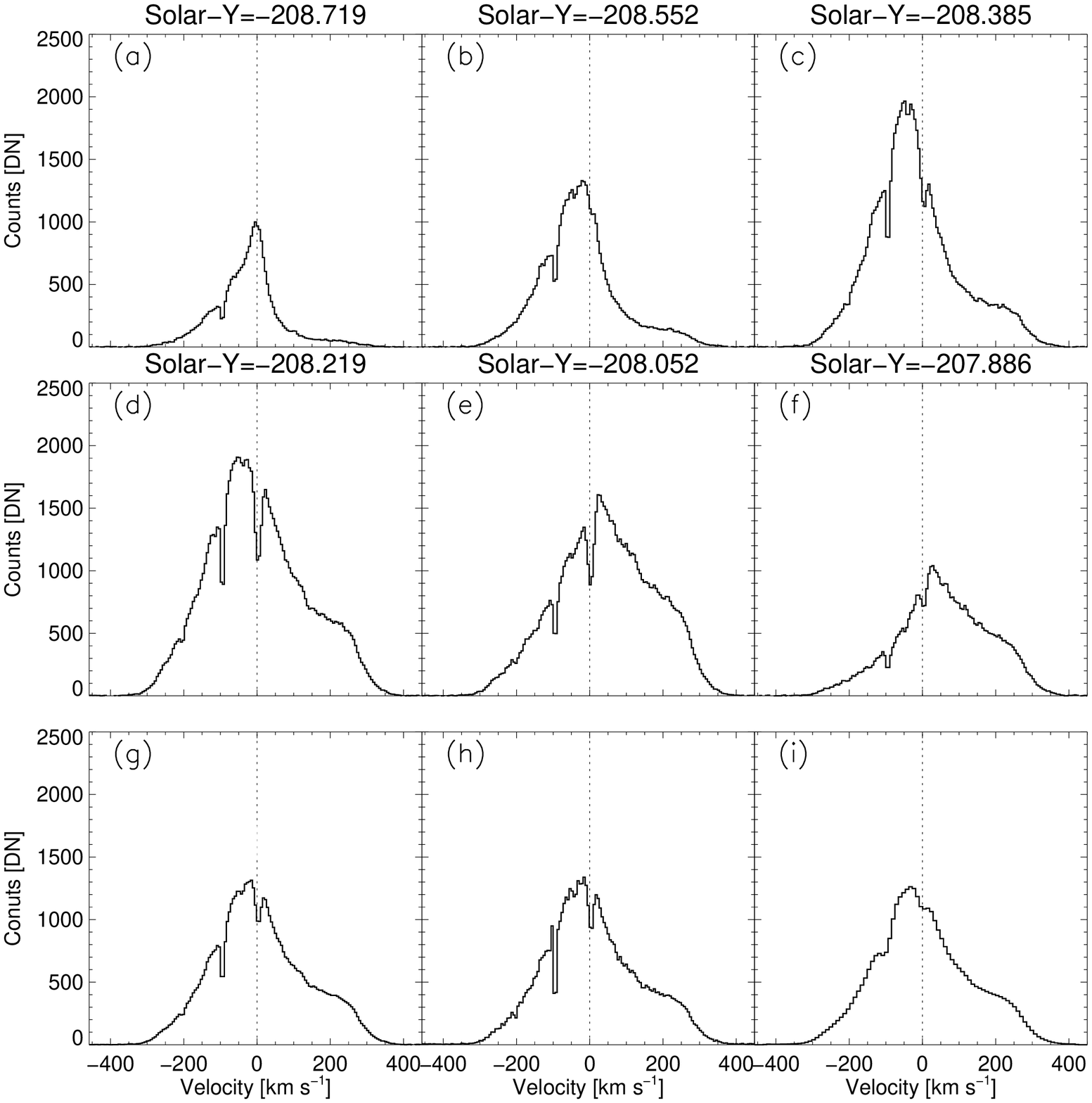}
\caption{Self-absorption of  Si\,{\sc{iv}} and its detectability with IRIS and\ SUMER.
\newline
(a) to (f): Si\,{\sc iv} 1394\,\AA\ line profiles obtained with IRIS at six adjacent pixels along the slit at 09:22:13~UT. (g): The average line profile of the six   Si\,{\sc iv} profiles observed with IRIS. (h): The average line profile {\itshape after}  the six individual profiles were deconvolved
in the spectral direction, now plotted in histogram mode to emphasize the spectral sampling. (i): Profile from panel (h) after the convolution with the SUMER instrumental profile and rebinned to the SUMER spectral sampling. This is what SUMER would have seen from this event. See Sect.\,\ref{S:sumer}.} \label{fig4}
\end{figure*}

\begin{figure*}

   \centering
    \includegraphics[width=7cm]{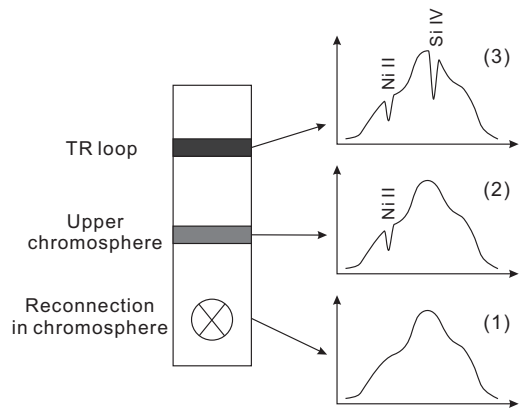}

\caption{Cartoon illustrating a scenario for the formation of the (self-)absorption features in Si\,{\sc iv}.
\newline
Below the upper chromosphere an reconnection event occurs leading to a significantly enhanced brightness and width of Si\,{\sc iv},  similar to what is shown in panel (1). On its way through the atmosphere the light first passes through the overlaying cool structures in the  upper chromosphere, where Ni\,{\sc ii} ions lead to absorbtion, as indicates in panel (2). Still higher up there is   Si\,{\sc iv}, e.g. in overlying cool loops,  that can lead to a narrow self-absorption feature.  This then leads to a spectra as illustrated in panel (3). See Sect.\,\ref{S:discussion}.}
      \label{fig5}
\end{figure*}

In contrast to \citet{Peter2014Sci} we also find self-absorption features in the Si\,{\sc{iv}} lines. We see two possible interpretations for this.
One option is that  the (pre-existing) upper atmosphere above the event contains a sufficient amount of Si\,{\sc{iv}} to cause the self-absorption (panel 3 of Figure\,\ref{fig5}). This is supported by the 1400\,\AA\ slit-jaw-images that show an indication of loops crossing the location on the slit at the location the event is observed (cf.\ Figure\,\ref{fig3}b). That is the case for the main event studied here, but even more so for the second case we briefly mention (cf. bottom red box in Figure\,\ref{fig3}b and profile in Figure\,\ref{fig3}c). In the compact brightening events reported
by
\cite{Peter2014Sci}  no clear
self-reversal can be identified from
the Si\,{\sc iv} line profiles. In their case no obvious Si\,{\sc iv} loop
structures are visible above the brightening event, which would be consistent with this interpretation.

Alternatively, the Si\,{\sc{iv}} self-absorbtion could also arise within the source region of the Si\,{\sc{iv}} brightening event itself, at the height of the middle chromosphere.
The decreased ratio of the Si\,{\sc{iv}} resonance lines (from about 2 for optically thin to 
\textbf{$1.77\pm0.05$}) speaks in favour of this interpretation. The spectra reported by \cite{Peter2014Sci} showed no self-absorption and had a ratio of the Si\,{\sc{iv}} lines always being close to 2. For a definite conclusion on the origin of the self-absorption feature of Si\,{\sc{iv}} further studies are needed, observational investigations as well as considerations including radiative transfer.

The seemingly common physical cause of  quite different phenomena, e.g. explosive events \citep[e.g.][]{Innes1997Natur}, Ellerman-bombs and related features
\citep{Peter2014Sci} and the event we report here, gives rise to an interesting question: What is the real difference between these events. Could it be that the magnetic changes underlying these phenomena
are similar, and that only the location of the energy deposition determines the different observational consequences? Extended studies incorporating IRIS and other instruments might bring us closer to the answer of this intriguing question.

\acknowledgments{IRIS is a NASA Small Explorer mission developed and operated by LMSAL, with mission operations
executed at NASA Ames Research Center and major contributions to downlink communications funded by the Norwegian Space Centre through a European Space Agency PRODEX contract. The Solar Dynamics Observatory (SDO) data that we used are provided courtesy of NASA/SDO and the Atmospheric Imaging Assembly and Helioseismic and Magnetic Imager science teams. We would like to thank Davina Innes, Pengfei Chen, Zongjun Ning, Leping Li, William Forman for helpful comments. The group from Peking University was supported by NSFC under 41174148, 41222032, 41231069, 41474147, 41274172, 41474148, and 41421003. Limei Yan was also supported by the short-term overseas project for doctoral student from Peking University. H.T. was supported under contract 8100002705 from LMSAL to the SAO. The work of F.C. and K.B. was supported by the International Max Planck Research School (IMPRS) for Solar System Science at the University of G\"ottingen.}

\newpage
\bibliographystyle{apj} 

\end{document}